\documentclass[prd,twocolumn,showpacs,preprintnumbers,tightenlines,amssymb,
               eqsecnum,superscriptaddress,floatfix]{revtex4}
\usepackage{epsfig,hyperref,dcolumn}

\begin{document}
\preprint{AEI-2003-006}
\preprint{NSF-KITP-03-55}

\title{A numerical relativistic model of a massive particle in orbit near a
       Schwarzschild black hole}

\author{Nigel T. Bishop}
\affiliation{Department of Mathematics, Applied Mathematics and
Astronomy, University of South Africa, P.O. Box 392, Pretoria
0003, South Africa}

\author{Roberto G\'omez}
\affiliation{Pittsburgh Supercomputing Center, 4400 Fifth Ave.,
Pittsburgh, PA 15213, U.S.A.} \affiliation{Department of Physics
and Astronomy, University of Pittsburgh, Pittsburgh, PA 15260, U.S.A.}

\author{Sascha Husa}
\affiliation{Max-Planck-Institut f\"ur Gravitationsphysik,
Albert-Einstein-Institut, Am M\"uhlenberg 1, D-14476, Golm, Germany}

\author{Luis Lehner}
\affiliation{Department of Physics and Astronomy, Louisiana State
University, Baton Rouge, LA 70810, U.S.A.} \affiliation{University
of British Columbia, Vancouver, BC, Canada V6T-1Z1}

\author{Jeffrey Winicour}
\affiliation{Max-Planck-Institut f\"ur Gravitationsphysik,
Albert-Einstein-Institut, Am M\"uhlenberg 1, D-14476, Golm,
Germany}
\affiliation{Department of Physics and Astronomy,
University of Pittsburgh, Pittsburgh, PA 15260, U.S.A.}

\date{September 10, 2003}

\begin{abstract}
We present a method for computing the evolution of a spacetime containing
a massive particle and a black hole. The essential idea is that the
gravitational field is evolved using full numerical relativity, with the
particle generating a non-zero source term in the Einstein equations.
The matter fields are not evolved by hydrodynamic equations. Instead
the particle is treated as a quasi-rigid body whose center follows a geodesic.
The necessary theoretical framework is developed and then implemented
in a computer code that uses the null-cone, or characteristic, formulation
of numerical relativity. The performance of the code is illustrated in test
runs, including a complete orbit (near $r=9M$) of a Schwarzschild black
hole.
\end{abstract}


\maketitle

\section{Introduction}

This paper is concerned with the evolution of a spacetime
containing a small object in orbit near a black hole, the goal being to
compute the motion of the object and the emitted gravitational
radiation. In the techniques normally used for this type of
`radiation-reaction' problems, the small
object is treated as a point-particle evolving on a fixed background spacetime with
its self-force taken into account. There are a number of approaches
concerning the implementation of the self-force -- for
example~\cite{mino2003,barack2003,blan,gual,bara,pfen,balo,lous,quinwald,
tanaka,hughes,burko2002}. An
alternative approach that could be used is full numerical relativity including
computational relativistic hydrodynamics (see for
example~\cite{fogo,shur,gour,fara,oona,philiptoni}, and for a
review~\cite{shib}). To our knowledge, such simulations have not been
performed for the extreme mass ratios considered in this paper.

The approach developed here uses full numerical relativity but with the
hydrodynamic aspect greatly simplified, and can be described as a {\em polytropic
particle (PP) method}. We use full numerical relativity for the evolution of the
gravitational field (including a non-zero stress energy tensor as source).
The matter fields are not evolved by relativistic hydrodynamics but rather
the object is treated as a quasi-rigid body whose center is evolved by the geodesic
equation. This approach avoids the intricacies and computational expense of
relativistic hydrodynamics.

Of course, there are limitations to the PP approach: it can only be
applied in situations in which the internal hydrodynamics of the material
object are unimportant, and in which the rigidity approximation is reasonable.
For example, one would certainly need full numerical relativity and relativistic
hydrodynamics in situations in which the object is expected to be tidally
disrupted. On the other hand, if tidal effects are small (in a sense that will
be made precise), then the PP model should provide a good description of
the physics.

The bounds of the domain of applicability of the PP method have not been
investigated systematically. However, from the
example runs presented in Sec.~\ref{sec:res}, the method should be
applicable under appropriate restrictions, e.g. when $m \leq 10^{-5} M$
where $M$ is the mass of the black hole and $m$ is the mass of the particle,
and when the size of the
particle is small enough that it not be tidally disrupted. Thus, we expect the
particle method to be applicable in astrophysical situations involving the
inspiral and capture of a neutron star or white dwarf by a galactic black hole
(with $M$ about $10^6 M_\odot$). On the other hand, it is difficult to see how
the method could be used for a stellar remnant black hole (with $M$ about $10
M_\odot$) -- any object with small enough $m/M$ would have too large a
diameter. Thus the method is expected to make predictions concerning
gravitational radiation that will be relevant to observations by
LISA~\cite{finn}, rather than to observations by LIGO or other earth based
detectors.

The results presented here use a characteristic gravity
code~\cite{hpn,roberto}. However, there is no reason to restrict the PP
method to the characteristic approach. The method should also be applicable
within Cauchy formulations of numerical relativity.

We have also constructed a finite difference version of a
$\delta-$function model. We found that the PP model performs better,
giving smoother results and, in particular, exhibiting convergence with
grid-refinement (as described in Sec.~\ref{sec-conv}). For these reasons,
although we describe both models, we give implementation details and results
only for the PP model.

The purpose of this paper is to introduce the PP method. We also give
some example runs exhibiting inspiral and a plunge to the black hole. These
runs demonstrate the potential of the method, and are not an
accurate description of the physics -- in particular, in the inspiral case,
the inspiral rate is much larger than that predicted by other methods.
Of course, validated physical results are
the goal of this project but, as discussed in Sec.~\ref{sec:con}, more
computational testing and development is needed before the goal can be attained.

We begin by summarizing previous results on the
characteristic formulation of numerical relativity in
Sec.~\ref{sec:nota}. Issues concerning the theoretical
framework of a massive particle are discussed in Sec.~\ref{sec:thry}.
Sec.~\ref{sec:com} presents, in detail, the computational
algorithms. Tests of the code and example runs
are given in Sec.~\ref{sec:res}.

\section{Summary of previous results, and notation}
\label{sec:nota}

The formalism for the numerical evolution of Einstein's equations,
in null cone coordinates, is well known~\cite{iww,hpn,cce} (see
also~\cite{ntb93,ntb90,rai83,bondi,tam}). For the sake of
completeness, we give here a summary of the formalism, including
some of the necessary equations. The version of the gravity code
being used here is fully described in~\cite{roberto}.

We use coordinates based upon a family of outgoing null hypersurfaces.
We let $u$ label these hypersurfaces, $x^A$ $(A=2,3)$, label
the null rays and $r$ be a surface area coordinate. In the resulting
$x^\alpha=(u,r,x^A)$ coordinates, the metric takes the Bondi-Sachs
form~\cite{bondi,sachs}
\begin{eqnarray}
   ds^2 & = & -\left(e^{2\beta}(1 + {W \over r}) -r^2h_{AB}U^AU^B\right)du^2
        - 2e^{2\beta}du dr  \nonumber \\ &-& 2r^2 h_{AB}U^Bdudx^A
         + r^2h_{AB}dx^Adx^B,
\label{eq:bmet}
\end{eqnarray}
where $h^{AB}h_{BC}=\delta^A_C$ and
$det(h_{AB})=det(q_{AB})$, with $q_{AB}$ a unit sphere metric.
We work in stereographic coordinates $x^A=(q,p)$ for which the unit sphere
metric is
\begin{equation}
q_{AB} dx^A dx^B = \frac{4}{F^2}(dq^2+dp^2),
\mbox{ where }
        F=1+q^2+p^2.
\end{equation}
(In previous notation we used $P=1+q^2+p^2$. Here we change notation to $F$
because $P$ now represents pressure, which we cannot denote by $p$ because that
is a stereographic coordinate). We also introduce a complex dyad
$q^A=\frac{F}{2}(1,i)$ with $i=\sqrt{-1}$. For an arbitrary Bondi-Sachs metric,
$h_{AB}$ can then be represented by its dyad component
\begin{equation}
J=h_{AB}q^Aq^B/2,
\end{equation}
with the spherically symmetric case characterized by $J=0$.
We introduce the (complex differential) eth operators $\eth$ and $\bar \eth$
(see~\cite{eth} for full details), as well as a number of auxiliary variables
$K=h_{AB}q^A \bar q^B /2$, $U=U^Aq_A$, $Q_A=r^2 e^{-2\,\beta}h_{AB}U^B_{,r}$,
$Q=Q_Aq^A$, $B=\eth\beta$, $\nu=\bar\eth J$ and $k=\eth K$.

The Einstein equations decompose into hypersurface equations, evolution
equations and conservation laws. The hypersurface equations form a hierarchical
set for $\nu_{,r}$, $k_{,r}$, $\beta_{,r}$, $B_{,r}$, $(r^2 Q)_{,r}$, $U_{,r}$
and $W_{,r}$; and the evolution equation is an expression for $(rJ)_{,ur}$. The
explicit form of the equations is given in~\cite{roberto} in the vacuum case.
The matter source terms, in the case of a perfect fluid of density $\rho$, pressure
$P$ and velocity $v^\alpha$, are stated in~\cite{mat}; except the matter
source term in Eq.~\cite{mat}--(31) is incorrect and that equation should read
\begin{eqnarray}
    && 2 \left(rJ\right)_{,ur}
    - \left((1+r^{-1}W)\left(rJ\right)_{,r}\right)_{,r} =
    -r^{-1} \left(r^2\eth U\right)_{,r} \nonumber \\
    &&+ 2 r^{-1} e^{\beta} \eth^2 e^{\beta}- \left(r^{-1} W \right)_{,r} J
    + N_J \nonumber \\
    && +\frac{4 e^{2\beta}\pi(\rho + P)}{r}
    \left( (J \bar{V}_{ang} - K V_{ang})^2 +V^2_{ang}   \right),
\label{eq:Ju}
\end{eqnarray}
where $V_{ang}=v_A q^A$~\cite{mat}, and $N_J$ is defined
in~\cite{hpn} and~\cite{roberto}. The remaining Einstein equations reduce to
conservation conditions which need only be satisfied on the inner boundary,
which are automatically satisfied here because the boundary has a simple
Schwarzschild geometry.

The null cone problem is normally formulated in the region of spacetime between
a timelike or null worldtube $\Gamma$ and ${\cal I}^+$. We represent
${\cal I}^+$ on a finite grid by using a  compactified radial coordinate
$x=r/(1+r)$. The numerical grid is regular in $(x,q,p)$ and consists of two
patches (north and south), each containing $n_x n_q n_p$ grid-points. The
$x-$grid covers the range $[0.5, 1]$. Each angular grid patch extends
two grid-points beyond the domain $(q,p)\in [-q_s,q_s]\times[-q_s,q_s]$, with
$q_s \ge 1$. Thus there is an overlap region at the equator with larger
overlap for larger $q_s$.

We denote the Bondi-Sachs metric (\ref{eq:bmet}) by $g_{\alpha\beta}$ and the
background metric ($g_{\alpha\beta}$ with $J=U=\beta=0$, $W=-2 M/r$) by
$g_{[M]\alpha\beta}$. The mass $M$ of the black hole is normally scaled to $M=1$
in simulations.

\section{Theoretical framework}
\label{sec:thry}

We have developed two different particle models with rather different
conceptual frameworks but implemented with very similar numerical codes. This
section describes each of the two frameworks, as well as some other theoretical
issues.

\subsection{The polytropic particle (PP) model}
\label{sec:poly}

The PP model treats the particle as an object of fixed
size in its local proper rest frame, with its center $z^\alpha (u)$ describing a
geodesic of the full spacetime. The particle is treated as quasi-rigid, and
a simple formula is used to evaluate the stress-energy tensor, which then appears as
source in the Einstein equations.
The simplest model of a polytrope, which will be used here, is for the case
with index $n=1$. Then the density $\rho$ and pressure $P$ of an
equilibrium configuration are given by~\cite{chandra}
\begin{equation}
\rho=\frac{m \sin{\frac{R \pi}{R_*}}}{4 R R_*^2},
        \;\; P=\frac{2 R_*^2 \rho^2}{\pi},
\label{eq-pl1}
\end{equation}
for $R \le R_*$, and $\rho=P=0$ for $R > R_*$, where $R_*$ is the radius of the
polytrope and $R$ is the distance from its center.

The density $\rho(R)$ and pressure $P(r)$ at a point $x^\alpha=(u,x^i)$ in the
Bondi-Sachs coordinate system are set by determining the distance $R$, in the
local proper rest frame, between $x^\alpha$ and the geodesic described by the
center of the polytrope. We first define the displacement vector
$\epsilon^\alpha$ relative to the polytrope's center,
\begin{equation}
\epsilon^u = 0, \;\; \epsilon^i = x^i - z^i(u).
\label{eq-pl2}
\end{equation}
The projection of $\epsilon^\alpha$ into the hypersurface orthogonal to
the worldline at time $u$ is
\begin{equation}
R_\alpha = (g_{\alpha\beta} + v_\alpha v_\beta) \epsilon^\beta,
\label{eq-pl3}
\end{equation}
with $g_{\alpha\beta}$ evaluated at $z^\alpha$, and $v_\alpha$ evaluated at
time $u$. Then we define the proper distance $R$ as the magnitude of $R_\alpha$,
\begin{equation}
R = \sqrt{g^{\alpha\beta} R_\alpha R_\beta},
\label{eq-pl4}
\end{equation}
and set the density and pressure using Eq. (\ref{eq-pl1}).
Thus, in its local proper rest frame, a (small) polytropic particle is a spherically symmetric
object with proper radius $R_*$.
The perfect fluid condition is used to set the stress-energy tensor, i.e.
$T_{\alpha \beta}=(\rho + P) v_\alpha v_\beta + P g_{\alpha \beta}$.

The PP model is approximate in the following sense.
Equation~(\ref{eq-pl1}) is exact only for an isolated sphere in equilibrium in
Newtonian theory. In general relativity it is a good approximation if
$m \ll R_*$ (and the example runs presented later satisfy this condition).
Further, in the field of a black hole, a polytrope would not
preserve its spherical shape but would become tidally distorted. Both these
approximations introduce the same type of error -- the PP's motion is
treated as quasi-rigid but, for the purpose of determining its
gravitational field, its stress-energy tensor $T_{\alpha\beta}$ is modeled as a
perfect fluid. Thus there are
contributions to $T_{\alpha\beta}$ that are being ignored, and the
magnitude of these contributions is now estimated.

Firstly, since the simulations in Sec.~\ref{sec:res} satisfy the condition $m \ll R_*$,
the error in ignoring the tidal stress will dominate that of using Eq.~(\ref{eq-pl1})
to estimate the pressure.
We use Newtonian theory to make an order of magnitude estimate of the
tidal stress. The tidal acceleration in the radial direction is $ 2M x/r^3$
where $x$ is the distance from the center of the polytrope. The tidal stress is
maximum at the center of the polytrope, and a simple dimensional argument
(which can be confirmed by integration) shows that
\begin{equation}
\sigma_{\mbox{max}}={\mathcal O} \left( \frac{Mm}{ r^3 R_*} \right) .
\end{equation}
In order to estimate the significance of  $\sigma_{\mbox{max}}$ in the
stress-energy tensor, we compare it to the maximum density
\begin{equation}
\frac{\sigma_{\mbox{max}}}{\rho_{\mbox{max} }}\equiv S
  ={\mathcal O} \left( \frac{4 R^2_*M}{\pi r^3} \right).
\end{equation}
In a physical situation in which the polytrope is stable against tidal
disruption, $(M R_*^3)/(m r^3) < {\mathcal O}(1)$, and the ratio $S$ is smaller than
${\mathcal O}(m/R_*)$, which, as previously noted, is small here.
However, in the example to be considered in Sec.~\ref{sec:orbit}, the
polytrope is not stable against tidal disruption, and thus the internal
stresses may be large. In fact, the ratio $S$ is approximately
$4/(81 \pi) \approx 0.016$.

\subsection{The $\delta-$function model}
\label{sec:del}

At the analytic level, a point particle of mass $m$
at position $z^\alpha=(u,z^i)=(u,r,z^A)$ has density $\rho$
and 4-velocity $v^\alpha$ satisfying
\begin{equation}
  \rho\sqrt{-g} v^u = m\delta(x^i-z^i),
\end{equation}
where $\int \delta(x^i-z^i) dr dq dp =1$. We model the $\delta$-function on the
grid by assigning weights $w$ to each grid point in a stencil surrounding
the particle. In terms of a test function $\phi$, this requires
\begin{equation}
  \phi(z^i)=\Sigma_I \phi_I w_I \Delta_V
\end{equation}
where $\Sigma_I$ is a sum over a stencil $I$ of grid points
surrounding the particle position $z^i$ and
$\Delta_V$ is the coordinate 3-volume of the stencil $I$.
We determine the weights $w_I$ representing the
$\delta$-function by choosing a set of test
functions, e.g. for the stencil of eight points determined
by the cell surrounding the particle we choose
\begin{eqnarray}
    \phi&=&a+a_i(x^i-z^i)+a_{ij}(x^i-z^i)(x^j-z^j)\nonumber\\
     &+&a_{ijk}(x^i-z^i)(x^j-z^j)(x^k-z^k),
\end{eqnarray}
where $i\ne j\ne k$ so that the $a$'s constitute eight arbitrary coefficients.
This then gives 8 simultaneous equations to solve for the $w_I$, which are
given explicitly in Eq. (\ref{eq-wts}) below.

It is necessary to renormalize the metric so as
to avoid infinities in the equations of motion. The metric occurs through the
normalization of the 4-velocity $v_\alpha$ and the raising of
indices. We take the components $(v_r,v_A)$ to be basic since they represent
the pullback of the 4-velocity to the null hypersurface. We renormalize the
other components by using the background metric $g_{[M]\alpha\beta}$ to raise
indices and to normalize the 4-velocity. This avoids the problem of an infinite
self-potential energy of the particle and is in keeping with the principle
that the energy of the particle only depends on its velocity and position in
the Schwarzschild field. It should be emphasized that this renormalization, or use of
$g_{[M]\alpha\beta}$ rather than $g_{\alpha\beta}$, applies only
to the {\em undifferentiated} metric. Metric derivatives that
occur in the particle equations of motion are computed using the
full metric $g_{\alpha\beta}$ -- otherwise radiation reaction could not be
included and we would simply be computing the motion of a test particle in the
Schwarzschild geometry. Of course, it is the full metric which is evolved by
the characteristic algorithm.

Although we were able to use the $\delta$-function model to compute qualitatively
reasonable orbits for the problems considered in Secs.~\ref{sec:orbit}
and \ref{sec:infall}, they were significantly less smooth
than with the PP model, and the growth in the deviation from
a Schwarzschild orbit was much faster. Further, we
did not find any quantity that exhibited convergence,
making it problematic to use the $\delta$-function model to obtain physical
predictions. Perhaps a more sophisticated numerical approach might
lead to convergence of global quantities, but this would be a difficult project
that we do not pursue here.

\subsection{Modeling the particle orbit}
\label{sec-appc}

A goal of this work is to study radiation reaction. This is a small effect,
and, numerically, it could be hidden if terms of order unity (representing the
background Schwarzschild geometry) are added and subtracted in the
equations of motion.
The motion of a test particle in a background Schwarzschild geometry satisfies
certain conservation laws; the motion of a particle with mass  $0<m\ll M$ does not
satisfy these laws, but they are nevertheless useful because these laws
indicate quantities that change very slowly. Our strategy is to use the
Schwarzschild conservation laws to find quantities that, in the general case,
evolve slowly. In the process, all background Schwarzschild terms cancel out
and we are left with expressions involving only small quantities.

For the case of a test particle in the Schwarzschild geometry
there is a reflection symmetry plane, the plane of the orbit. Thus
the normalized 4-velocity is completely determined by its
components $T^\alpha v_\alpha =v_u$ and $\Phi^\alpha v_\alpha
=v_\phi$, where $T^\alpha$ and $\Phi^\alpha$ are the Killing
vectors of the Schwarzschild background and $v_\phi$ is a
velocity component with respect to $(u,r,\theta,\phi)$
null-spherical coordinates. In the general case, $v_u$ and
$v_\phi$ are approximately conserved.
Partly because of the stereographic coordinates being used, the
implementation is quite technical (see Sec.~\ref{sec-impc} for details).

\subsection{Caustics}

The characteristic evolution code breaks down if caustics develop, which render
the null coordinate system employed singular. A rough estimate can be readily
obtained by employing the well-known condition for the deflection of light by a
massive body such as the Sun. We find as an approximate condition for caustics
not to form that
\begin{equation}
\frac{R_*^2}{4 m} > r.
\label{eq:caust}
\end{equation}

\section{Computational method}
\label{sec:com}

\subsection{Overview}

The PP method evolves both the matter and gravity fields.
At each grid-point at which the density is non-zero, the particle's
density, pressure and 3-velocity are found and used to construct the
right hand side of the Einstein equations,
and the gravitational field is then evolved as described in
Sec.~\ref{sec:nota}.
The gravitational field affects the motion of the particle: the 3-velocity
$v_i$ is evolved by using the geodesic equation in the form
\begin{equation}
\frac{d v_i}{d u} = \frac{\Gamma_{\alpha i \gamma} v_\delta v_\epsilon
                     g^{\alpha\delta} g^{\gamma\epsilon}}{v^u};
\label{eq-vi}
\end{equation}
and the particle's position is evolved by
\begin{equation}
\frac{d z^i}{d u}=\frac{v^i}{v^u}.
\label{eq-z}
\end{equation}

The setting of initial data is described in Sec.~\ref{sec:ini} below. The
worldtube $\Gamma$ at $r=2M$ is the (past) horizon of a Schwarzschild black
hole of mass $M$. Thus the boundary data on $\Gamma$ has the simple analytic
form~\cite{mat}
\begin{equation}
J=\nu=k=\beta=B=U=Q=0,\;\; W=-2M.
\end{equation}

\subsection{Computational algorithms}

The iterative evolution algorithm proceeds as follows:

\begin{enumerate}
\item Start at time $u=u^{(n)}$. The gravitational field
$g^{(n-1)}_{\alpha\beta}$ is known over the whole grid and the boundary data
supplies $g^{(n)}_{\alpha\beta}$ in a neighborhood of $r=2M$. The particle's
position $z^{i(n)}$ and velocity $v_i^{(n)}$ are also known.
\item Determine the grid-cell $G^{(n)}_P$ containing the point $z^{i(n)}$;
i.e., determine $a_i$ such that
\begin{eqnarray}
 r^{(a_1)} < z^{1(n)} < r^{(a_1+1)}, \nonumber \\
 q^{(a_2)} < z^{2(n)} < q^{(a_2+1)}, \nonumber \\
 p^{(a_3)} < z^{3(n)} < p^{(a_3+1)}.
 \end{eqnarray}
This is done on both north and south patches, although if the particle is not
in the equatorial overlap region there will be a solution for only one of the
patches.  We define
\begin{eqnarray}
 \Delta^{i(n)} = x^{a_i+1}-x^{a_i}, \quad && \delta_0^{i(n)}=x^{a_i+1} - z^{i(n)},
      \nonumber \\ && \delta_1^{i(n)}=z^{i(n)}-x^{a_i}
\end{eqnarray}
and we define weights at the eight grid-points at the corners of $G^{(n)}_P$ by
\begin{equation}
w(x^{a_i+e_i})=\frac{\delta^{1(n)}_{e_1}\delta^{2(n)}_{e_2}\delta^{3(n)}_{e_3}}
               {\Delta^{1(n)}\Delta^{2(n)}\Delta^{3(n)}} ,
\label{eq-wts}
\end{equation}
where $e_i =\{0,1\}$.
\item Next, we set the density and pressure. In general, this needs to be done
on both north and south patches. The density at the grid-point $x^i$ is set by
means of Eqs.~(\ref{eq-pl1})-(\ref{eq-pl4}); then the pressure $P$ is set.
\item The Einstein equations are now integrated to find the metric
$g_{\alpha\beta}^{(n)}$. The source terms are given in~\cite{mat} (where, as
already noted, Eq.~\cite{mat}-(31) should be replaced by Eq.~(\ref{eq:Ju})).
\item The formula $g^{\alpha \beta} v_\alpha v_\beta =-1$ is used to find
$v^{(n)}_u$; the metric $g^{\alpha \beta}$ is known at the required
grid-points and its value at the particle position $z^{i(n)}$ is found by
taking a weighted average using the weights found in Eq. ( \ref{eq-wts}) above.
\item The formula $v^\alpha = g^{\alpha \beta} v_\beta$ is used to
find $v^{\alpha (n)}$, again
using the weighted average to find $g^{\alpha \beta}$.
\item Equation (\ref{eq-z}) is now used to find $z^{i(n+1)}$. On the first
time-step, this is done by the Euler method and, on subsequent time steps,
by the 3-point Adams-Bashforth method, i.e.
\begin{equation}
z^{(n+1)}=z^{(n)}+\frac{\Delta u}{2} \bigg( 3\frac{dz}{du}^{(n)}
                            - \frac{dz}{du}^{(n-1)} \bigg)
\end{equation}
\label{en-z}
\item We now find $v_i^{(n+1)}$. The right hand side of Eq.~(\ref{eq-vi})
is evaluated at $z^{i(n)}$ by, as usual, finding the value at the grid-points
and taking a weighted average using the weights found in Eq. ( \ref{eq-wts}) above.
The terms in Eq.~(\ref{eq-vi}) are quite complicated and were found using a
Maple script, which was also used to generate Fortran code. Details are given
in an Appendix. The numerical evolution method is the same as used in
step~\ref{en-z}.
\end{enumerate}

\subsection{Setting the initial data}
\label{sec:ini}

We have experimented with various ways of setting the initial data, and found that,
apart from some early transient effects, the options tried make little difference to both
the particle orbit and the gravitational field.
The initial gravitational content is prescribed by setting $J=0$ at $u=0$. (It is
also possible to set the initial data $J$ by a Newtonian limit
condition~\cite{newt1,newt2,iww}, the computational implementation of which
will be discussed elsewhere). Of course, this means that we are introducing
spurious gravitational radiation into the initial data, which might have the effect of
initializing the particle into a different orbit to that intended. We tried evolving the
code for a pre-determined time $u_S$ (and $u_S$ could be possibly set to zero),
during which time the particle's velocity and position are not updated.
The idea is that the gravitational field should relax to the
correct form, with the spurious initial gravitation content radiating away,
by the time $u_S$ when the particle is allowed to move.

The code requires the initial velocity as a 1-form $v_i$ but a physical
description normally specifies the tangent vector $v^i$. For example, a
particle in a circular orbit would have
\begin{equation}
v^r=0, \;\; (v^p)^2+(v^q)^2 = \frac{F^2 M}{4r^2(r-3 M)}.
\end{equation}
Suppose that we are given $v^i$ rather than $v_i$. Initially, when only
the background metric is known, $v_i$ is constructed from $v^i$ using
\begin{equation}
g_{[M]\alpha\beta} v^\alpha v^\beta =-1
\end{equation}
to first determine $v^u$; then $v_i=g_{[M]i\alpha} v^\alpha$.
Then, while $u \le u_S$, the code uses the
fact that $v^\alpha$ is found at each time step to determine a value of $v_r$
such that $v^r=0$ by an iterative algorithm. Explicitly, we use the secant
algorithm
\begin{equation}
v_{r(a+1)}=v_{r(a)} - v^r_{(a)}f_S\frac{v_{r(a)}-v_{r(a-1)}}{v^r_{(a)}-v^r_{(a-1)}},
\end{equation}
where $a$ is the iteration number and $f_S$ is a factor (which is 1 in the
standard algorithm) that may need to be set to 0.2 or smaller for stable
convergence -- the difficulty here is that we are solving $v^r(v_r)=0$ not as a
simple algebraic equation but as an equation whose coefficients change as the
metric relaxes.

We found that the value of $u_S$ has little effect on computed orbits, at least
for the cases computed in Sec.~\ref{sec:orbit} below. Thus, we will set
$u_S=0$. Nevertheless, the option of setting a non-zero $u_S$ is retained in
the code in case circumstances are found in which a smoother evolution is obtained.

\subsection {Implementation of the approximate conservation laws}
\label{sec-impc}

The theoretical basis for using approximate angular momentum and energy conservation
to improve the accuracy of the orbit computation,
was discussed in Sec.~\ref{sec-appc}. We now present details of how this is
implemented for (1) the angular momentum in an equatorial orbit, (2) the
angular momentum in a polar orbit, and (3) the energy. The code is written so
that all of these approximate conservation laws may be used, or not, simply by
changing input parameter switches.

\subsubsection{Angular velocity in an equatorial orbit}

The angular momentum per unit mass
\begin{equation}
     h= q v_p-p v_q
\label{eq-h}
\end{equation}
is approximately conserved. In terms of proper time $\tau$ along the particle's
trajectory,
\begin{equation}
   \frac{dh}{d\tau} =
   v^q v_p + q \frac{dv_p}{d\tau} -v^p v_q - p \frac{dv_q}{d\tau}.
\label{eq-dh}
\end{equation}
Now, Eq. (\ref{eq-vi}) takes the form
\begin{equation}
\frac{d}{d\tau}v_A = -\frac{z^A}{r^2} \left( (v_p)^2 + (v_q)^2 \right) + E_A, \;\;
 A=(q,p),
\label{eq-dva}
\end{equation}
where the $E_A$ contain only small quantities. We also introduce the small
quantity $\mu^i=(g^{i\alpha}-g_{[M]}^{i\alpha})v_\alpha$, which represents
the difference between raising an index of the covariant velocity by the full
or background metric. Thus,
\begin{equation}
    v^A=\mu^A + \frac{v_A}{r^2}.
\label{eq-mua}
\end{equation}
Combining Eqs.~(\ref{eq-dh}), (\ref{eq-dva}) and (\ref{eq-mua}), we obtain
\begin{equation}
\frac{dh}{d\tau} = \mu^q v_p  -\mu^p v_q + q E_p - p E_q,
\label{eq-dh1}
\end{equation}
which is implemented in the code. We extract $v_A$ from the evolved value of
$h$. This is done by using the constraint that the particle is on the equator,
$q^2+p^2=1$. Thus $q v^q + p v^p =0$ so that
\begin{equation}
 q \left(\mu^q + \frac{v_q}{r^2}\right)
                               + p\left(\mu^p + \frac{v_p}{r^2}\right) =0.
\label{eq-heq}
\end{equation}
Combining Eqs. (\ref{eq-h}) and (\ref{eq-heq}), we find
\begin{eqnarray}
  v_q &=& - p h -r^2 q (q \mu^q +p \mu^p) \nonumber \\
   v_p &=& q h -r^2 p (q \mu^q +p \mu^p),
\label{eq-hqp}
\end{eqnarray}
which is implemented in the code. Furthermore, the particle is constrained to
follow the equator exactly, and so the particle's position is corrected
according to $z^A \rightarrow z^A f_e$ with
\begin{equation}
f_e = \sqrt{\frac{1}{q^2+p^2} }.
\end{equation}

\subsubsection{Angular velocity in a polar orbit}

In the case of polar motion, simplified here to the case $p=0$, the equations
analogous to Eqs. (\ref{eq-h}), (\ref{eq-dh1}) and (\ref{eq-hqp}) are
\begin{equation}
h= \frac{F v_q}{2}, \;\; \frac{dh}{d\tau}=\frac{F A_q}{2} + q v_q \mu^q,
                   \;\; v_q=\frac{2h}{F}.
\label{eq-pole}
\end{equation}

\subsubsection {The energy}

The energy per unit mass $v_u$ is conserved along a geodesic in the Schwarzschild
background. In this
case
\begin{equation}
v_{uS} = \frac{h^2+r^2+v_r^2 (r^2-2Mr)}{2 v_r r^2}.
\label{eq-v0}
\end{equation}
We take $v_{uS}$, as defined above, to be an approximately
conserved quantity. From Eq.~(\ref{eq-vi}),
\begin{equation}
\frac{d}{d\tau}v_r = -\frac{h^2}{r^3}  -\frac{v^2_r M}{r^2} + E_1.
\label{eq-dv1}
\end{equation}
Using
\begin{equation}
\frac{dr}{d\tau}=v^r= -v_{uS}+\left(1-\frac{2M}{r}\right) v_r +\mu^r,
\label{eq-mu1}
\end{equation}
differentiation of Eq. (\ref{eq-v0}) leads to
\begin{eqnarray}
   \frac{d}{d\tau}v_{uS} = \Big( 2 h
   \frac{dh}{d\tau}rv_r+2v_r^3Mr\mu^r-2h^2v_r\mu^r \nonumber
   \\ +E_1(r^3 v_r^2-r^3-2Mr^2v_r^2-h^2r) \Big) \frac{1}{2r^3v_r^2}.
\label{eq-dv0}
\end{eqnarray}
There is an option in the code to evolve $v_{uS}$ by Eq. (\ref{eq-dv0}). In
this case, we extract $v_r$ from the value of $v_{uS}$. This is done by
rewriting Eq.~(\ref{eq-v0}) as a quadratic in $v_r$. We find
\begin{equation}
    v_r= \frac{v_{uS}\pm \sqrt{v_{uS}^2 - \left( 1-\frac{2M}{r} \right)
    \left(1+\frac{h^2}{r^2} \right) }}{\left( 1-\frac{2M}{r} \right)} .
\label{eq-v02v1}
\end{equation}
When the code is evolving $v_{uS}$ by Eq.~(\ref{eq-dv0}), at each
time step it also evolves $v_r$ in the usual way. The $\pm$ in
Eq.~(\ref{eq-v02v1}) is chosen so that the result for $v_r$ is
closest to the directly evolved value; further, if the square root
in Eq.~(\ref{eq-v02v1}) is less than some threshold, or imaginary,
the directly evolved value of $v_r$ is not corrected. For a
circular orbit of the Schwarzschild background, the square root is
exactly zero, and therefore it is difficult to use this option
when evolving a circular orbit.

\subsection{The metric variable $W$}

The only Bondi-Sachs metric variable that is non-zero in the background
metric is $W$, and in order to improve numerical accuracy, the code treats
$W$ as the sum of the background analytic part
($W_{\mbox{an}}$) plus a correction ($W_{\mbox{num}}$). The values of
$W_{\mbox{an}}$ and its derivatives are found exactly, and finite differencing
is applied only to the part $W_{\mbox{num}}$. In effect, this also applies to
the other metric variables, because their background analytic parts are zero.

\section{Computational tests and results}
\label{sec:res}

\subsection{Convergence}
\label{sec-conv}

\subsubsection{Initial accelerations}
Firstly, we investigated the convergence of various accelerations
on the initial null cone, and in so doing tested the gravitational hypersurface
equations, the gravitational evolution equation and the particle evolution
equations. The tests were made with the particle initialized at $r=9$ at the
north pole with $v^r=v^p=0$ and $v^q$ set to the value for a circular orbit. The
particle mass was $m=10^{-4}$. The particle velocity was updated directly,
without incorporating approximate
conservation laws. The overlap between north and south patches was minimal
($q_s=1.0$). The size of the polytrope was $R_*=5.0$. The following
quantities, all of which are rates of change, were determined on the initial
null cone: $||J_{,u}||_\infty$, $h_{,u}$ $v_{u,u}$, $v_{r,\tau}$ and
$v_{q,\tau}$. The quantities involving $u-$derivatives were found by evolving
the code for one time-step and then applying the formula
$Q_{,u}=(Q_1-Q_0)/\Delta u$; the quantities involving $\tau-$derivatives are
found directly by the code using data only on the initial null cone.

The following grids were used: (a)
coarse, $n_x=41$, $n_q=n_p=25$; (b) medium $n_x=81$, $n_q=n_p=45$;
and (c) fine, $n_x=161$, $n_q=n_p=85$. In the different grids,
$\Delta_x$ and $\Delta_q=\Delta_p$ scale as 4:2:1. The (single) time-step was
$\Delta u=10^{-5}$, which, for all grids, is much smaller than the spatial
discretization, so that second order spatial accuracy is expected.
Assuming  that a quantity $Q$ behaves as $Q=a+b\Delta^n$, it
is  straightforward to show that
\begin{equation}
n= \log_2 \frac{Q_c-Q_m}{Q_m-Q_f}
\label{eq:convrate}
\end{equation}
where $Q_c$, $Q_m$ and $Q_f$ refer to the computed values of $Q$
using the coarse, medium and fine grids, respectively.

Our results are stated in Table~\ref{tab:poly}: it is clear that, on the
initial null cone, the polytropic model is convergent with the order $n$
in the range 1.59 to 2.28.
\begin{table}[!b]
\caption{Convergence of the Polytropic model\label{tab:t1}}
\begin{ruledtabular}
\begin{tabular}{lrrrd}
                    & Coarse     & Medium     & Fine       & $n$  \\
$||J_{,u}||_\infty$ &  0.4346$\times 10^{-2}$ &  0.4441$\times 10^{-2}$ &  0.4460$\times 10^{-2}$ & 2.28 \\
$h_{,u}$            & -0.7377$\times 10^{-2}$ & -0.3436$\times 10^{-2}$ & -0.2129$\times 10^{-2}$ & 1.59 \\
$v_{u,u}$           &  0.2732$\times 10^{-4}$ &  0.1273$\times 10^{-4}$ &
0.0789$\times 10^{-4}$ & 1.59 \\
$v_{r,\tau}$        & -0.5050$\times
10^{-3}$ & -0.5455$\times 10^{-3}$ & -0.5564$\times 10^{-3}$ & 1.9  \\
$v_{q,\tau}$        & -1.8069$\times 10^{-3}$ & -0.8416$\times 10^{-3}$ &
-0.5215$\times 10^{-3}$ & 1.59  \end{tabular}  \end{ruledtabular}
\label{tab:poly}
\end{table}

\subsubsection{Circular orbit}
\label{s:con-o}
Secondly, we performed a convergence test for a particle in a circular orbit around
a black hole. For the coarse and medium grids, the particle completed a whole orbit,
but for the fine grid this was not possible.
The particle was initialized at $r=9$, $q=0$ and $p=1$ with $v^r=v^p=0$
and $v^q$ set to the value required for a circular equatorial orbit. The mass of the
particle was $m=10^{-6}$ and the size was $R_*=3$ (the requirement that the
polytrope should be resolvable by all grids places a lower limit on $R_*$).
We used the technique in Sec.~\ref{sec-impc} to model approximate conservation of
angular momentum, but not of energy. The angular
grid-patch overlap was $q_s=1.2$. The test results are for the time interval
$0 \le u \le 20$ representing just under $1/8$ of a whole orbit. The grids used were
(a) coarse, $n_x=61$, $n_q=n_p=20$ with $du = 1.6666 \times 10^{-2}$; (b) medium,
$n_x=121$,  $n_q=n_p=35$, $du=8.3333 \times 10^{-3}$; and (c) fine, $n_x=241$,
$n_q=n_p=65$, with $du=4.1666 \times 10^{-3}$.

\begin{figure}[!t]
\epsfig{file=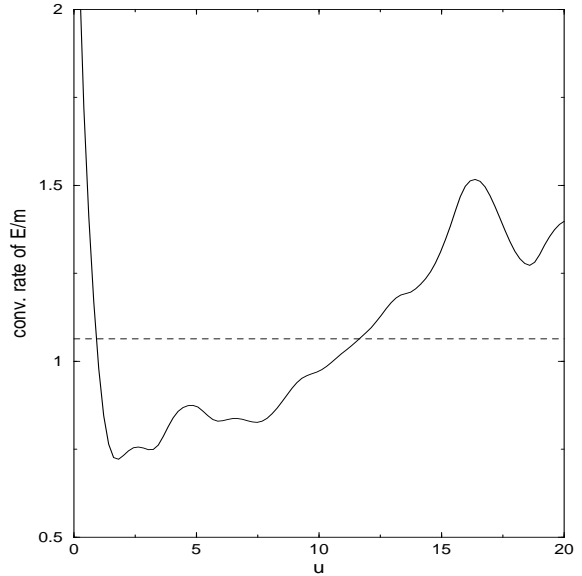,height=3.5in,width=3.5in,angle=0}
\caption{Convergence rate of $E/m$ as a function of time for $0<u<20$.}
\label{f:v0_conv}
\end{figure}

The convergence rate of $v_u$, between $u=0$ and $u=20$ as
estimated from Eq.~(\ref{eq:convrate}), is shown in Fig.~\ref{f:v0_conv};
the average rate is $n=1.064$, so that the effective
convergence rate of the particle's energy is approximately first order.
This is an artifact of the numerical scheme used in the evolution
equation, which for a fixed value of the dissipation parameter is only
first order in time~\cite{hpn}.

\subsection{Whole orbit with $m \ne 0$}
\label{sec:orbit}

The various input parameters were described in Sec.~\ref{s:con-o} ({\it
circular orbit}), medium grid. The computation was run for 25,000 time steps
until $u=208$ and represents more than one orbit (which is achieved at about
$u=170$); the computation took about 24 hours of wall-clock
computer time. As discussed in
Sec.~\ref{s:con-o}, the results are within the convergence
regime of the numerical method.
 The run was performed for illustrative purposes and is not
physical because a polytrope with the parameters used here would be tidally
disrupted.

\begin{figure}[!t]
\epsfig{file=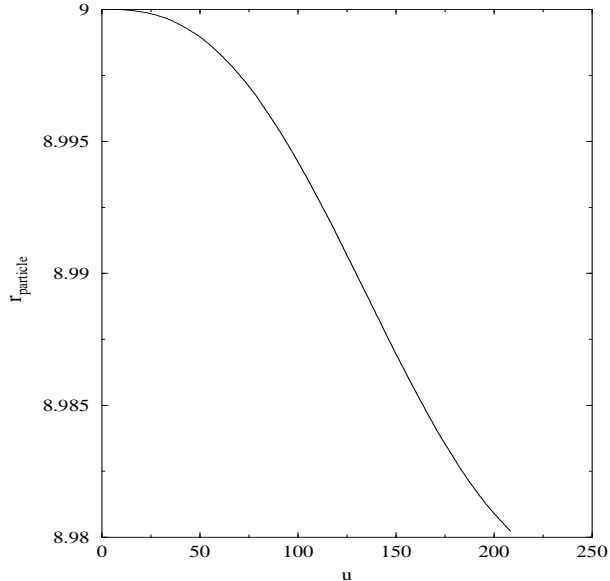,height=3.5in,width=3.5in,angle=0}
\caption{$r$-coordinate for a complete orbit.}
\label{f:o-r}
\end{figure}

\begin{figure}[!t]
\epsfig{file=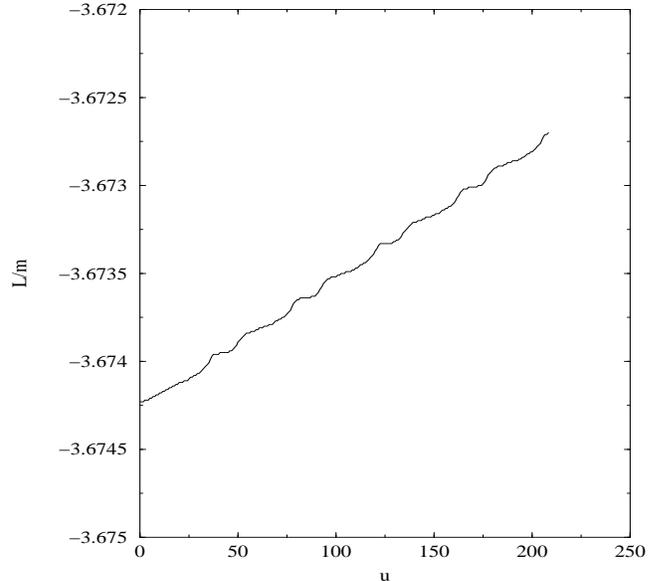,height=3.5in,width=3.5in,angle=0}
\caption{Angular momentum per unit particle mass for a complete orbit.}
 \label{f:o-h}
\end{figure}

\begin{figure}[!t]
\epsfig{file=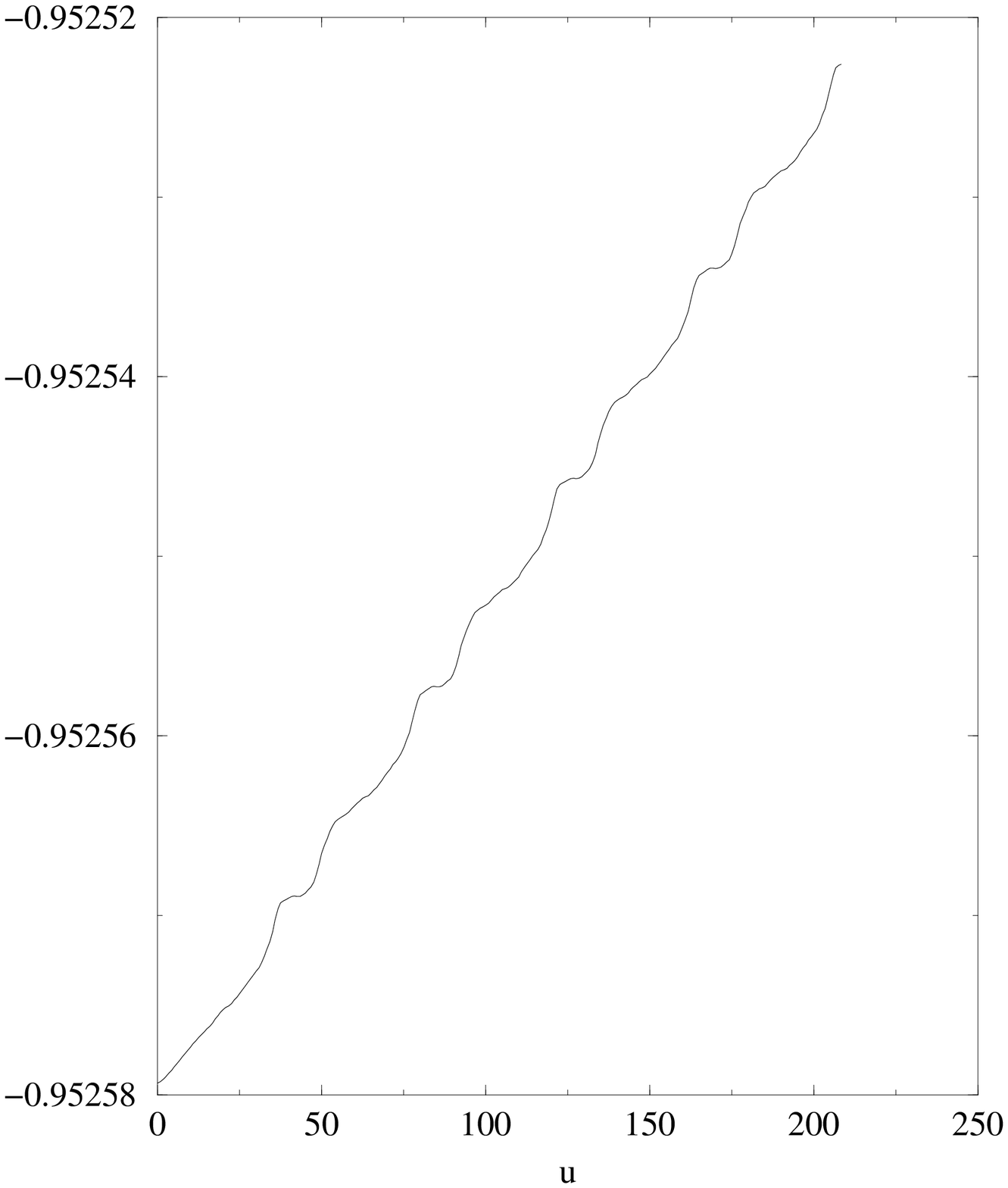,height=3.5in,width=3.5in,angle=0}
\caption{Energy per unit particle mass for a complete orbit.}
\label{f:o-v0}
\end{figure}

\begin{figure}[!t]
\epsfig{file=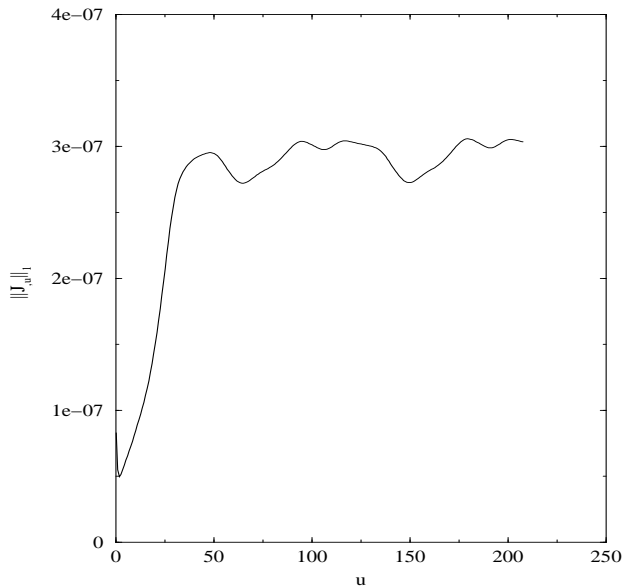,height=3.5in,width=3.5in,angle=0}
\caption{$L_1$ norm of the radiation indicator, $\parallel J_{,u}\parallel_1$, for a complete orbit.}
 \label{f:o-j}
\end{figure}

The results of the computation are shown in Figs.~\ref{f:o-r} to
\ref{f:o-j}, in which
the particle inspirals (Fig.~\ref{f:o-r}), losing energy (Fig.~\ref{f:o-v0}) and angular
momentum (Fig.~\ref{f:o-h}).
Figure~\ref{f:o-j} shows the time development of the $L_1$ norm of $J_{,u}$,
which provides a measure of the dynamic activity of the gravitational field.
We also performed whole orbit computations with input parameters as
above, but varying the particle mass by powers of 10 in the range $10^{-4}$ to
$10^{-9}$. We found that the energy loss rate scales, as expected, as $m^2$.

In our numerical method, the measured inspiral was $\Delta r=-0.016$ after
one complete orbit. The inspiral after one orbit is
$\Delta r=-2.98 \times 10^{-6}$ according to the quadrupole formula~\cite{MTW}, and
$\Delta r=-4.75 \times 10^{-6}$ according to a perturbative method~\cite{hughes,sah}.
The rate of energy loss, i.e. the rate of
change of $E= m v_u$, is a better measure of the inspiral rate -- because $\Delta r$
after one orbit includes a contribution due to the orbit becoming slightly
elliptical. Averaged over a
complete orbit, the measured rate for the numerical method is
$dE/du= 2.669 \times 10^{-13}$, whereas the quadrupole formula predicts
$dE/du= 1.10 \times 10^{-16}$, and the perturbative method gives
$dE/du= 1.75 \times 10^{-16}$.
There is thus a discrepancy between the energy loss rates found here and
by other methods. The cause of the discrepancy is not known, and may comprise a
number of factors: (a) in order to
resolve the particle properly, we are forced to make its size too large for it to be physical,
and thus the model ignores internal tidal stresses that in this case are large (see
Sec.~\ref{sec:poly}); (b) lack of resolution; and (c) other. The issue is discussed further
in the Conclusion, Sec.~\ref{sec:con}.

\subsection{Capture of particle by the black hole}
\label{sec:infall}

The purpose of the test was to see how the code behaves as the particle
approaches the event horizon at $r=2$, and the particle was initialized at
$r=6$, i.e. at the ISCO. The size of the particle was $R_*=2$, and so, as
with the complete orbit computation, the model ignores significant tidal
stresses and is not physical. In order to shorten the inspiral time
the particle was given a small inward radial velocity ($v^r=-0.01$), and the
angular velocity was set to that for a circular orbit in the test particle
limit ($v^q=0.0962$, $v^p=0$). The test was performed with two different
grids, enabling us to see at which stage numerical errors become significant.
The grids used were $n_x=121$, $n_q=n_p=35$, $du=8.3333 \times 10^{-3}$
(medium grid); and $n_x=81$, $n_q=n_p=25$, $du=1.25 \times 10^{-2}$
(coarser grid).

\begin{figure}[!b]
\epsfig{file=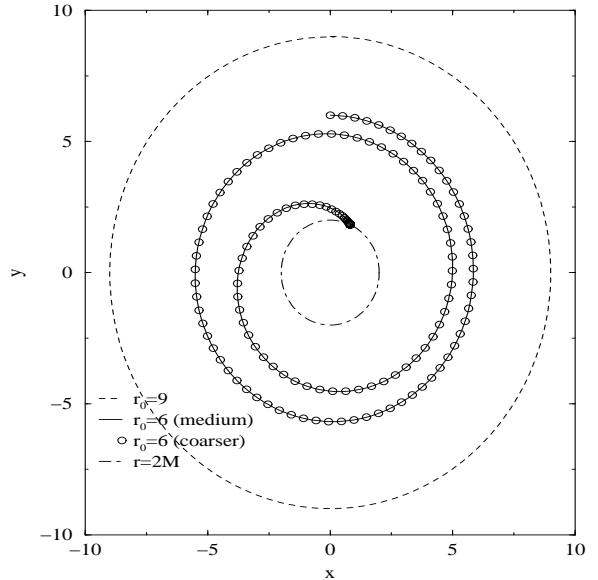,height=3.48in,width=3.5in,angle=0}
\caption{The orbit, in the $(x,y)=(r\cos \phi,r \sin \phi)$ plane,
traced by a particle, initially at $r=6$, as it is captured by a
black hole. Overlaid in the figure is the orbit traced by a
particle initially at $r=9$ (dotted line). The central circle
indicates the location of the horizon ($r=2$). }
 \label{f:h-p-r}
\end{figure}

\begin{figure}[!b]
\epsfig{file=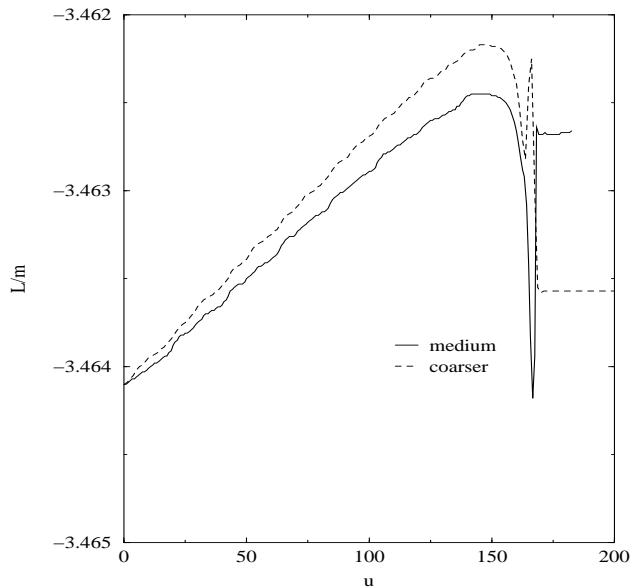,height=3.5in,width=3.5in,angle=0}
\caption{Angular momentum per unit particle mass for the capture
of a particle by a black hole.} \label{f:h-p-h}
\end{figure}

\begin{figure}[!b]
\epsfig{file=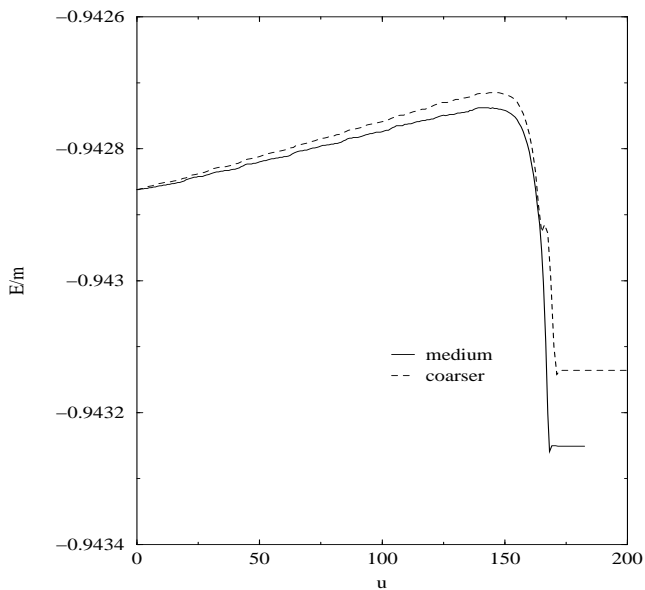,height=3.5in,width=3.5in,angle=0}
\caption{Energy per unit particle mass for the capture of a
particle by a black hole.} \label{f:h-p-v0}
\end{figure}

\begin{figure}[!t]
\epsfig{file=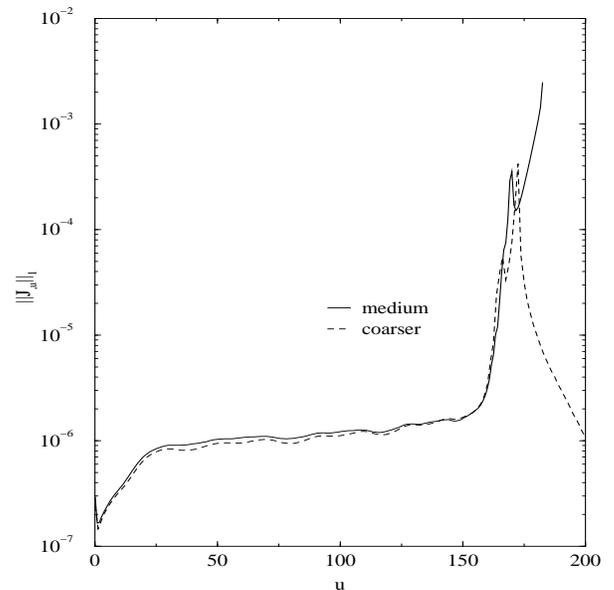,height=3.5in,width=3.5in,angle=0}
\caption{$L_1$ norm of the radiation indicator, $\parallel J_{,u}\parallel_1$,
shown using a logarithmic scale in the vertical axis, for the capture of a
particle by a black hole.} \label{f:h-p-j}
\end{figure}

In the coordinates being used, as $r \rightarrow 2$ the evolution
variable $v_r\rightarrow\infty$. Thus, because of this coordinate effect, we
expect the code to crash at some value of $r$ just greater than $2$.
The results of the computation are shown
in Figs.~\ref{f:h-p-r} to \ref{f:h-p-j}.
In the medium grid computation, the particle inspirals until the code crashes at $u=182.5$
with the particle at $r=2.00077$ and $|v_r|\approx 5,000$. The particle
completed just over two complete revolutions, i.e. its angular position
changed by just over $4 \pi$ radians during the evolution. The particle
crossed $r=2.1$ at $u=162.5$, $r=2.01$ at $u=172.5$ and $r=2.001$ at $u=181.7$;
thus demonstrating a freezing of radial position, as expected due to the
redshift inherent in the $u$-coordinate. Throughout the computation, the
position of the particle varies smoothly (Fig.~\ref{f:h-p-r}). The particle loses energy
(Fig.~\ref{f:h-p-v0}) and
angular momentum (Fig.~\ref{f:h-p-h}) at a fairly constant rate, until about $u=150$, $r=3.2$.
Further, the activity of the gravitational field (as measured by $\parallel
J_{,u} \parallel$, Fig.~\ref{f:h-p-j}) starts to grow rapidly at this time. We have not analyzed
the cause of this effect. The results of the coarser grid match those of the
medium grid reasonably well until about $u=170$, at which stage $r\approx 2.02$.

\subsection{Other experiments and computational issues}

A run was performed using the same input parameters as in Sec.~\ref{sec:orbit},
except that $m=0$. The orbit
was found to be exactly circular, with no loss of energy or angular momentum.
Because the numerical method implements the approximate conservation of
angular momentum, which in this case is exact, the angular momentum behaved
as expected. However, conservation of
energy was not enforced. The only (minor) error was in the time
for an orbit: the numerical method yielded $\Delta u =169.638$ (for the grid used
in this specific run), whereas the
analytic value is $2 \pi r^\frac{3}{2} = 169.646$. In order to check that the exact
conservation of energy was not just a  consequence of the
orbit being exactly circular,
we also did a run for a non-circular orbit, starting at $r=9$, $q=0$ , $p=1$ and with
initial velocity $v^r=0$, $v^q=0.047$, $v^p=0$. The particle was found to move in an
orbit between $r=9$ and $r=11.79$, and there was a very small change in $v_u$:
energy was conserved, in the sense of $d v_u/du=0$, to the order of one part in
$10^{14}$.

A simple test of the code is to determine if it produces results concerning
caustic formation that are consistent with the estimate given in
Eq. (\ref{eq:caust}). We performed a number of short runs (up to 100 iterations),
each with a different value of particle mass $m$, and made a binary chop
search to find the critical value of $m$ above which the code crashes due to the
onset of caustic formation
(indicated by the metric variable $\beta \rightarrow\infty$ at ${\mathcal I^+}$). The
test was performed
with initial velocity $v^i=0$, initial position $r=9$ at the north
pole, and polytropic
radius $R_*=2$.  The grid discretization was $n_x=121$,
$n_q=n_p=35$, with time step $dt=8.3333\times 10^{-3}$.
We found that the code behaved properly with
$m=0.03$ but crashed when $m=0.04$, consistent with the critical value of
$m=0.111$ indicated by Eq. (\ref{eq:caust}).

Unfortunately, it is not possible, at this time, to present results on gravitational
radiation output. The module used in the code for calculating the
news~\cite{hpn} was originally developed and tested under conditions in which
the fields are well resolved at ${\mathcal I^+}$, which is not the case
for particle applications. Improvements in
the news computation have recently been investigated~\cite{yosefthesis, nnews}, but
it is not yet known whether the radiation from a particle source can be reliably
computed. Results will be reported elsewhere after the necessary development
and testing.

The tests were performed on a Linux machine with a single
processor running at 1.8GHz. The complete orbit run reported in Sec.~\ref{sec:orbit}
used a grid of $121  \times 35^2$ points (on each angular patch) and 25,000 time
steps. The run time was about 24 hours. Of course, the run time
scales with grid discretization as $\Delta^{-4}$.

\section{Conclusion}
\label{sec:con}

In this paper we have described and implemented the PP method for evolving the full
Einstein equations using a characteristic evolution code. The method could be adapted
and used within other numerical relativity frameworks. The PP method can be
a useful tool in modeling astrophysical situations involving a
black hole and another much smaller object, in a regime in which a full hydrodynamic
model is not necessary. We have demonstrated that the
method works in the sense that it computes orbits that are convergent, and deviations
from a Schwarzschild orbit scale as expected with particle mass $m$.
However, the computed inspiral rate was much larger than that predicted by other methods.
A feature of the code is that it avoids the computational expense of modeling
the material object by means of relativistic hydrodynamics.

The PP method has the potential to supply accurate orbits,
including inspirals towards the ISCO and plunges to the black hole, as well
as the associated gravitational radiation output. Such results would be useful both
directly, and also indirectly in providing error bars against which other methods
could be tested. In the tests described here, the need to resolve the particle has
prevented us from making the particle radius small enough so that its local rigidity
is justified. In order to achieve a proper physical basis for the model,
we envisage the following future work
\begin{enumerate}
\item It is necessary to investigate the effect of the polytrope radius ($R_*$) on
particle motion. On physical grounds, one would expect that in some regime
the particle motion should be
independent of $R_*$. (Of course, taking the limit $R_* \rightarrow 0$ is
equivalent to changing from the polytropic to a $\delta-$function model,
and this link provided motivation for the investigation of that model).
\item Once parameters for the model that are physically realistic can be attained,
it will be necessary to investigate whether the PP method computes
reliable energy loss rates.
\item The gravitational radiation output (Bondi news function) is required,
both to supply a waveform and to check the energy balance (the rate of loss of
orbital energy $m v_u$ should be of the same magnitude as the radiation power).
\item Once the above issues have been resolved, it will be necessary to
validate the PP method by obtaining results
that agree, in some regime, with results obtained by another method.
\end{enumerate}

\begin{acknowledgments}

We thank Manuela Campanelli and Carlos Lousto for discussions,
and Scott Hughes for providing information on perturbative results.
We benefited from the hospitality of the Max-Planck-Institut f\"ur
GravitationsPhysik, Albert-Einstein-Institut, the Center for Gravitational
Wave Physics at Pennsylvania State University, and the Kavli Institute
for Theoretical Physics at the University of California at Santa Barbara.
This work was partially supported by the National Research Foundation, South
Africa under Grant number 2053724, and by NSF Grants PHY-9988663 to the
University of Pittsburgh and PHY-0135390 to Carnegie Mellon University,
and by NSF Grants PHY-0114375 to the Pennsylvania State
University, and PHY-9907949 to the University of California at Santa
Barbara.

\end{acknowledgments}

\appendix

\section{The geodesic equation}

We have used Maple to compute the form of Eq.~(\ref{eq-vi}) for
the metric~(\ref{eq:bmet}). The angular part
of $v_i$, $v_A$ is represented by the spin weighted quantity
$V_{ang}=v_A q^A$. Further, for ease of application to the
approximate conservation method (Secs.~\ref{sec-appc} and
\ref{sec-impc}), the formulas are presented with the zeroth order
quantities (in each case, the first line) shown separately from
the perturbative ($E_i$) quantities.

\begin{widetext}
\begin{eqnarray}
\frac{dv_r}{d\tau} &=& \frac{2\bar{V}_{ang} V_{ang} -
    v_r^2 (r W_{,r} -W) r }{2 r^3} \nonumber \\
     &+&\bigg(+ 4 \bar{V}_{ang} V_{ang}(1-K)
     - 2 (e^{-2 \beta}-1) v_r^2 (r W_{,r} -W) r  K
     - 2 J \bar{V}_{ang}^2  K
     + 4 \bar{V}_{ang} V_{ang} \bar{J} J
     - r \bar{V}_{ang} V_{ang} J_{,r} \bar{J} \nonumber  \\*
     &-& r \bar{V}_{ang} V_{ang} J \bar{J}_{,r}
     - 4 \beta_{,r} e^{-2 \beta} v_r r^3  \bar{U} V_{ang} K
     - 4 \beta_{,r} e^{-2 \beta} v_r r^3  U \bar{V}_{ang} K
     - 2 \bar{J} V_{ang}^2  K   \nonumber \\
     &+& 4 e^{-2 \beta} v_r^2  \beta_{,r} (r+W) r^2  K
     - 8 \beta_{,r} e^{-2 \beta} v_r r^3  v_u K
     + r J_{,r} \bar{V}_{ang}^2  K
     + 2 e^{-2 \beta} v_r r^3  \bar{U}_{,r} V_{ang} K \nonumber  \\*
     &+& 2 e^{-2 \beta} v_r r^3  U_{,r} \bar{V}_{ang} K
     + r \bar{J}_{,r} V_{ang}^2  K \bigg)  \frac{1}{4 K r^3 }; \label{e:v1} \\*
   \frac{dV_{ang}}{d\tau} &=&
 - \frac{(q +i p) \bar{V}_{ang} V_{ang}}{r^2} \nonumber \\
&+&  \bigg(       - 4 (q+ip) \bar{V}_{ang} V_{ang} (K-1)
     + 4 (\eth K) J \bar{J} \bar{V}_{ang} V_{ang}
     - 2 (\eth J) K \bar{J} \bar{V}_{ang} V_{ang}
     - 2 (\eth \bar{J}) K J \bar{V}_{ang} V_{ang} \nonumber \\*
     &-& 4 e^{-2 \beta} v_r (\eth \beta) r^2  U \bar{V}_{ang}
     - 4 e^{-2 \beta} v_r (\eth \beta) r^2  \bar{U} V_{ang}
     - 2 e^{-2 \beta} v_r^2  (\eth W) r
     + 4 e^{-2 \beta} v_r^2  (r+W) (\eth \beta) r \nonumber \\*
     &-& 8 e^{-2 \beta} v_r (\eth \beta) r^2  v_u
     + (\eth \bar{J}) J^2  \bar{V}_{ang}^2
     + (\eth J) \bar{J}^2  V_{ang}^2
     - 2 (\eth K) J K \bar{V}_{ang}^2
     - 2 (\eth K) \bar{J} K V_{ang}^2
     + 2 (\eth K) \bar{V}_{ang} V_{ang} \nonumber \\*
     &+& (\eth J) \bar{V}_{ang}^2  \bar{J} J
     + (\eth \bar{J}) V_{ang}^2  \bar{J} J
     + 4 i p \bar{J} V_{ang}^2
     + (\eth J) \bar{V}_{ang}^2
     + (\eth \bar{J}) V_{ang}^2
     + 4 q \bar{J} V_{ang}^2
     + 2 r^2  e^{-2 \beta} v_r (\eth \bar{U}) V_{ang} \nonumber \\*
     &+& 2 r^2  e^{-2 \beta} v_r (\eth U) \bar{V}_{ang}
     + 4 r^2  e^{-2 \beta} v_r q \bar{U} V_{ang}
     + 4 i r^2  e^{-2 \beta} v_r p \bar{U} V_{ang} \bigg)  \frac{1}{4 r^2}.
\label{e:vc}
\end{eqnarray}
\end{widetext}

\bibliography{particle3}

\end{document}